\documentclass[%
 reprint,
 amsmath,amssymb,
 aps,
]{revtex4-2}

\usepackage{graphicx}
\usepackage{dcolumn}
\usepackage{bm}
\usepackage{physics}
\usepackage{amsmath}
\usepackage{natbib}
\usepackage{booktabs}
\usepackage{array}
\usepackage{float}
\usepackage{multirow}
\usepackage[colorlinks,linkcolor=blue,urlcolor=blue,anchorcolor=blue,citecolor=blue]{hyperref}
\usepackage{placeins}

\begin{document}
	
	
	\title{Intrinsic Localized Excitons in MoSe$_2$/CrSBr Heterostructure}
	
	
    \author{Xinyue Huang$^{1,2}$}
    \author{Zhigang Song$^{3}$}
    \author{Yuchen Gao$^{1}$}
    \author{Pingfan Gu$^{1}$}
    \author{Kenji Watanabe$^{4}$}
    \author{Takashi Taniguchi$^{5}$}
    \author{Shiqi Yang$^{1}$}
    \email{yang$_$shiqi@pku.edu.cn}
    \author{Zuxin Chen$^{6}$}
    \email{chenzuxin@m.scnu.edu.cn}
    \author{Yu Ye$^{1,7,8}$}
    \email{ye$_$yu@pku.edu.cn}
    
\affiliation{%
$^{1}$State Key Laboratory for Artificial Microstructure $\rm{\&}$ Mesoscopic Physics and Frontiers Science Center for Nano-Optoelectronics, School of Physics, Peking University, Beijing 100871, China\\
$^{2}$Academy for Advanced Interdisciplinary Studies, Peking University, Beijing 100871, China\\
$^{3}$John A. Paulson School of Engineering and Applied Sciences, Harvard University, Cambridge, Massachusetts 02138, United States\\
$^{4}$Research Center for Electronic and Optical Materials, National Institute for Materials Science, 1-1 Namiki, Tsukuba 305-0044, Japan\\
$^{5}$Research Center for Materials Nanoarchitectonics, National Institute for Materials Science, 1-1 Namiki, Tsukuba 305-0044, Japan\\
$^{6}$School of Semiconductor Science and Technology, South China Normal University, Foshan 528225, China\\
$^{7}$Yangtze Delta Institute of Optoelectronics, Peking University, Nantong 226010 Jiangsu, China\\
$^{8}$Liaoning Academy of Materials, Shenyang, 110167,  China\\
}
	%
	
	
 
\begin{abstract}

We present a comprehensive investigation of optical properties in MoSe$_2$/CrSBr heterostructures, unveiling the presence of localized excitons represented by a new emission feature, X$^*$. We demonstrate through temperature- and power-dependent photoluminescence spectroscopy that X$^*$ originates from excitons confined by intrinsic defects within the CrSBr layer. The valley polarization of X$^*$ and trion peaks displays opposite polarity under a magnetic field, which closely correlates with the magnetic order of CrSBr. This is attributed to spin-dependent charge transfer mechanisms across the heterointerface, supported by density functional theory calculations revealing a type-II band alignment and spin-polarized band structures. Furthermore, the strong in-plane anisotropy of CrSBr induces unique polarization-dependent responses in MoSe$_2$ emissions. Our study highlights the crucial role of defects in shaping excitonic properties. It offers valuable insights into spectral-resolved proximity effects in van der Waals heterostructures between semiconductor and magnet, contributing to advancing spintronic and valleytronic devices.

\end{abstract}

\maketitle


\section{\label{sec:level1}Introduction}

Van der Waals (vdW) heterostructures, consisting of vertically stacked layers of atomically thin two-dimensional (2D) materials, have emerged as versatile platforms for investigating unique physical properties with promising applications. The field has seen significant advancements since the discovery of 2D vdW magnets like Cr$_2$Ge$_2$Te$_6$\cite{gong2017discovery} and CrI$_3$\cite{huang2017layer} in 2017, particularly in magnetic vdW heterostructures. The combination of the remarkable optical properties of transition metal dichalcogenides (TMDs) with the inherent magnetism of 2D magnets presents exciting prospects for spintronic and valleytronic device development\cite{huang2020emergent}. In particular, the break of the time-reversal symmetry resulting from the interaction between TMD and adjacent magnetic layers has been observed to induce spontaneous valley splitting and increased valley polarization in heterostructures such as WSe$_2$/CrI$_3$\cite{zhong2017van,seyler2018valley,zhong2020layer} and MoSe$_2$/CrBr$_3$\cite{ciorciaro2020observation,lyons2020interplay}. Furthermore, asymmetric magnetic interactions have been detected in MoSe$_2$/CrBr$_3$ heterostructures\cite{choi2023asymmetric} through polarization-resolved reflection spectroscopy. 

Although the discovery of strong magnetic proximity effects at the TMD/magnetic material vdW heterointerfaces is undoubtedly a breakthrough, other intriguing properties related to rich exciton physics remain to be explored. The significant reduction in dielectric screening effects in monolayer TMD leads to the emergence of new excitonic species in various TMD-based heterostructures\cite{chernikov2014exciton}, including interlayer excitons\cite{rivera2015observation,jauregui2019electrical}, moiré excitons\cite{tran2019evidence,seyler2019signatures,alexeev2019resonantly,jin2019observation}, and localized excitons\cite{joshi2020localized,mahdikhanysarvejahany2022localized,fang2023localization}. However, studies on vdW magnetic heterostructures have predominantly focused on the excitonic valley properties of TMD, with limited investigation of the origin of the spectral features. Previous studies on TMD/2D magnet heterostructures have revealed interesting excitonic characteristics and new types of excitons, yet their physical origins have not been thoroughly discussed, often oversimplified as neutral excitons or trions. For instance, in the MoSe$_2$/Cr$_2$Ge$_2$Te$_6$ heterostructure\cite{zhang2022electrically}, significant broadening of excitonic peaks and a peak redshift up to 10 meV have been observed. Additionally, in two other vertically stacked heterostructures, MoSe$_2$/MnPSe$_3$ and MoSe$_2$/FePSe$_3$\cite{onga2020antiferromagnet}, the appearance of new peaks between the designated exciton and trion peaks has been noted, although their underlying nature remains elusive. Considering the susceptibility of TMD excitons to the electronic structure of neighboring materials, a greater emphasis on spectral features in heterostructures of TMD/2D magnets is warranted. Furthermore, a systematic study of excitonic physics in these systems is crucial for a comprehensive understanding of the proximity effects involving various excitonic species.

Here, we delve into the optical characteristics of a magnetic vdW heterostructure comprising monolayer MoSe$_2$ and antiferromagnetic CrSBr. The selection of CrSBr as the magnetic substrate stems from its notable physical properties, such as air stability\cite{telford2020layered,ye2022layer}, robust magnetically-coupled excitons\cite{wilson2021interlayer,dirnberger2023magneto,wang2023magnetically}, and highly anisotropic electronic structure\cite{wu2022quasi,klein2023bulk}. Our study uncovers the presence of localized excitons within the heterostructure. Through a systematic analysis of the power dependence and temperature-induced changes in the PL spectra, we validate that the localized excitons are confined by defect potentials originating from the underlying CrSBr layer. Furthermore, localized excitons demonstrate unique valley polarization characteristics distinct from trions, along with in-plane anisotropic behavior absent in single-layer MoSe$_2$. Our findings advance the understanding of exciton properties and underscore the significance of substrate defects in TMD/vdW magnetic heterostructures.

\begin{figure*}
	\centering
	\includegraphics[width=1\linewidth]{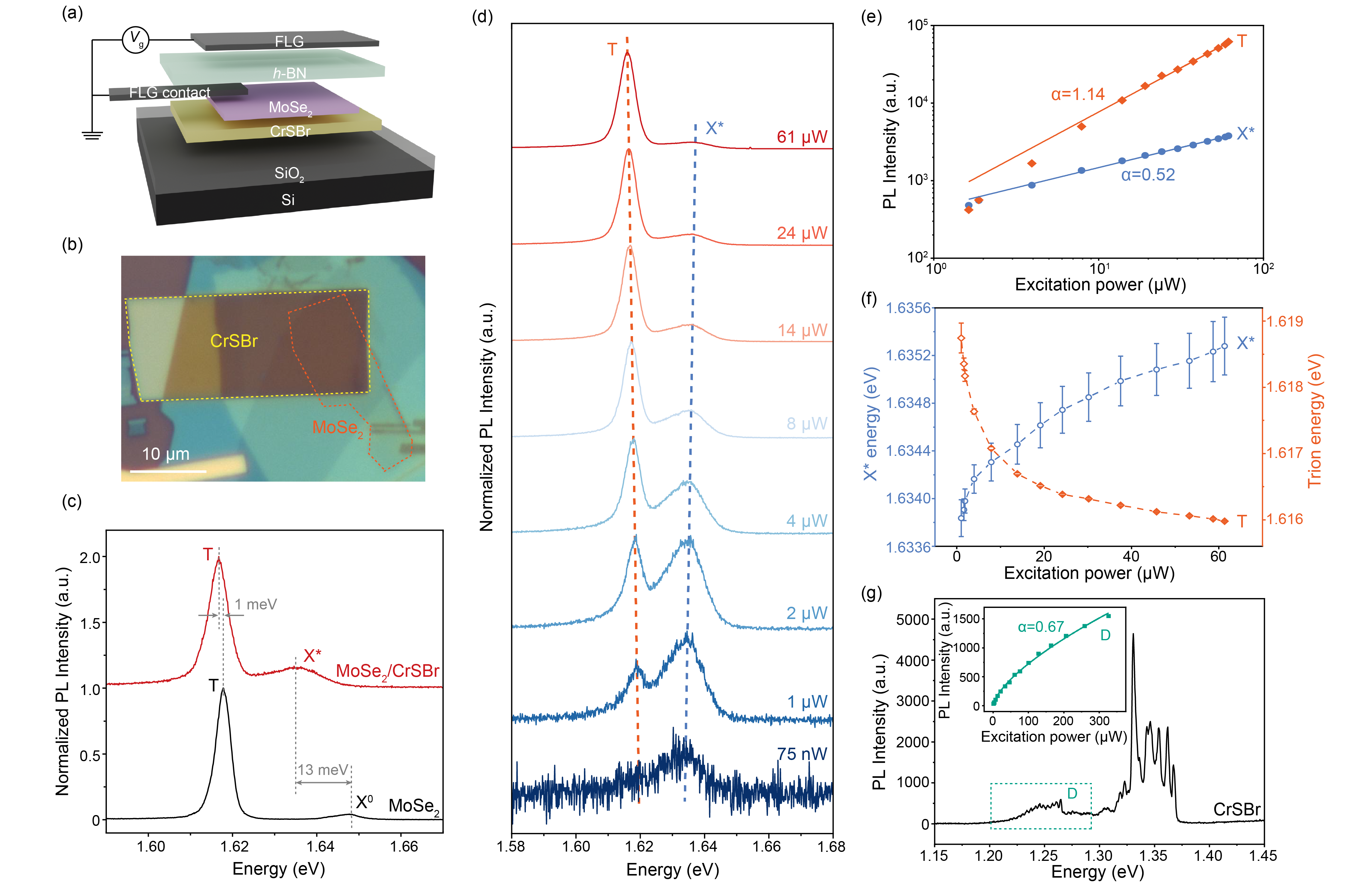}
	\caption{Basic characterization and power-dependent PL spectroscopy of MoSe$_2$/CrSBr heterostructure. (a) Schematic diagram of the device structure. (b) Optical image of device 1, with CrSBr and MoSe$_2$ outlined by yellow and orange dashed lines, respectively. The thickness of CrSBr flake is 47 nm. (c) PL spectra of monolayer MoSe$_2$ (black) and MoSe$_2$/CrSBr heterostructure (red). The T and X$^0$ peaks represent trion and neutral exciton, respectively. The new peak observed in the heterostructure is labeled X$^*$. (d) PL spectrum as a function of the excitation power. Each spectrum was individually normalized by its peak intensity. (e) Extracted PL intensities of trion and X$^*$ peaks as a function of excitation power in logarithmic coordinates. The solid lines are the power law fit to the data, and $\alpha$ represents the exponent of the fit. (f) Extracted X$^*$ peak (blue) and trion peak (orange) energies as a function of excitation power. (g) PL spectrum of CrSBr emissions in the heterostructure. The green dashed box represents the low-energy peaks emitted by optically active defects in the CrSBr substrate, denoted as D. The inset shows the power-dependent integrated intensity of the D peaks, showing a saturation behavior.}
	\label{Figure1}
\end{figure*}

\section{\label{sec:level2}Results and Discussions}

MoSe$_2$/CrSBr heterostructures were constructed using mechanical exfoliation and dry transfer techniques. A single-layer MoSe$_2$ was placed on top of a few-layer CrSBr flake, with a few-layer \textit{h}BN flake serving as the dielectric layer. Additionally, few-layer graphene flakes were utilized as gate and contact electrodes, allowing for electrostatic control of the carrier concentration within the heterostructure (Fig. \ref{Figure1}(a)). This letter focuses primarily on two devices (47 nm and 10 nm CrSBr), both exhibiting similar behavior. The optical image of device 1 is shown in Fig. \ref{Figure1}(b). Unless stated otherwise, all measurements were carried out at 2 K.

Fig. \ref{Figure1}(c) presents the normalized PL spectra of monolayer MoSe$_2$ (black) and MoSe$_2$/CrSBr device 1 (red) under 1.96 eV excitation energy. Since no obvious changes occur in the PL spectrum within the CrSBr emission window (see Supplemental Material Fig. S1), our attention is focused on the spectral features of MoSe$_2$ emissions. In the case of monolayer MoSe$_2$, the presence of two distinct peaks can be attributed to neutral excitons (X$^0$) at 1.648 eV and charged trions (T) at 1.618 eV. The high ratio of trions to excitons $I_\text{T}/I_{\text{X}^0}$ is a consequence of the high doping level in the monolayer MoSe$_2$\cite{ross2013electrical}. Similarly, two well-defined peaks are observed in the MoSe$_2$/CrSBr heterostructure. The trion peak at 1.617 eV exhibits a minimal shift ($\sim$ 1 meV) compared to monolayer MoSe$_2$. Notably, a new peak at 1.635 eV (referred to as X$^*$) emerges, positioned 18 meV above the trion peak and displaying a 13 meV redshift relative to the neutral excitons (X$^0$) in monolayer MoSe$_2$. The feature is consistent across all five heterostructures fabricated in this study, spanning CrSBr thicknesses from 3 nm to 47 nm (see Supplemental Material Fig. S2 and Table S1). The spatial distribution of the PL intensity reveals a uniform X$^*$ emission throughout the heterostructure (see Supplemental Material Fig. S3).  

To explore the nature of the X$^*$ peak, we initially conducted power-dependent PL measurements on device 1. Fig. \ref{Figure1}(d) shows the evolution of the two peaks as the excitation power increases. Notably, the X$^*$ peak dominates the PL spectra at low excitation power levels ($<$ 2 $\mu$W) but gradually reaches saturation, while the trion intensity experiences rapid growth and eventually surpasses the X$^*$ peak as the power increases. By fitting the intensity ($I$) of the two peaks using the power law  $I \propto P^\alpha$, we derive $\alpha$ = 1.14 for the trion peak and $\alpha$ = 0.52 for the X$^*$ peak (Fig. 1(e)). As documented in previous studies, the PL intensity power dependence for free exciton (trion) emission is expected to be linear, with $\alpha$ approaching 1. The sublinear behavior exhibited by the X$^*$ peak starkly contrasts with the trion peak and neutral exciton X$^0$ in monolayer MoSe$_2$ (see Supplemental Material Fig. S4), characteristic of potential trapped localized excitons\cite{schmidt1992excitation}.

In addition to changes in peak intensity, the energy of the trion and X$^*$ peaks exhibits distinct behaviors. As depicted in Fig. \ref{Figure1}(f), within the same excitation power range, the trion peak undergoes a redshift of approximately 2.5 meV, whereas the X$^*$ peak experiences a blueshift around 1.5 meV. The redshift observed in the trion peak could potentially stem from laser-induced heating effects or carrier-induced band gap renormalization\cite{ye2019charge}. In the case of the localized exciton X$^*$, the energy blueshift with increasing exciton density may be attributed to either the heightened occupancy of higher energy states within the trapping potential or the intensified repulsive interaction between excitons\cite{fang2023localization}.

Now, we discuss the physical origin of the potentials that confine the X$^*$ excitons. Since the emission of X$^*$ is uniform and reproducible in other samples, we consider the localized excitons to be intrinsic. As shown in Fig. \ref{Figure1}(c), both the monolayer MoSe$_2$ and the heterostructure PL spectra exhibit sharp and clean emissions with a linewidth less than 6 meV, indicating high-quality samples. Thus, excluding exciton localization from charge impurities in the monolayer MoSe$_2$ is straightforward and will be confirmed by magnetic field-dependent characteristics. On the contrary, a series of low-energy peaks (D) are observed in the emission window of CrSBr, with energies about 100 meV lower than the CrSBr exciton (see Fig. \ref{Figure1}(g)). The integrated intensity of the D peaks also shows a sublinear power dependence, $\alpha=0.67$. The observed saturation behavior of the D peaks is consistent with previous reports, attributed to emission from optically active defects in CrSBr\cite{klein2022sensing,marques2023interplay}. Recently, the role of intrinsic defects in CrSBr has been widely discussed. Early magnetotransport measurements revealed a strong coupling between charge transport and magnetic defects\cite{telford2022coupling}. Several groups have observed a low temperature magnetic transition around 40 K\cite{telford2020layered,lee2021magnetic,lopez2022dynamic}, with one hypothesis suggesting that it belongs to a hidden magnetic order formed by defects in CrSBr, although the exact origin is still under debate\cite{long2023intrinsic}. Using scanning tunneling microscopy (STM) imaging with atomic resolution, single Br vacancies (V$_{Br}$) with a surface density of $\sim$ 5 $\times$ 10$^{12}$ cm$^{-2}$ were observed in CrSBr flakes \cite{klein2022sensing}. Based on the above discussion, we propose that the X$^*$ excitons are localized by defect potentials in nearby CrSBr. The abundance and ubiquity of defects in CrSBr contribute to the homogeneous emission of MoSe$_2$ localized excitons and the complete quenching of neutral excitons.

\begin{figure}
	\centering
	\includegraphics[width=1\linewidth]{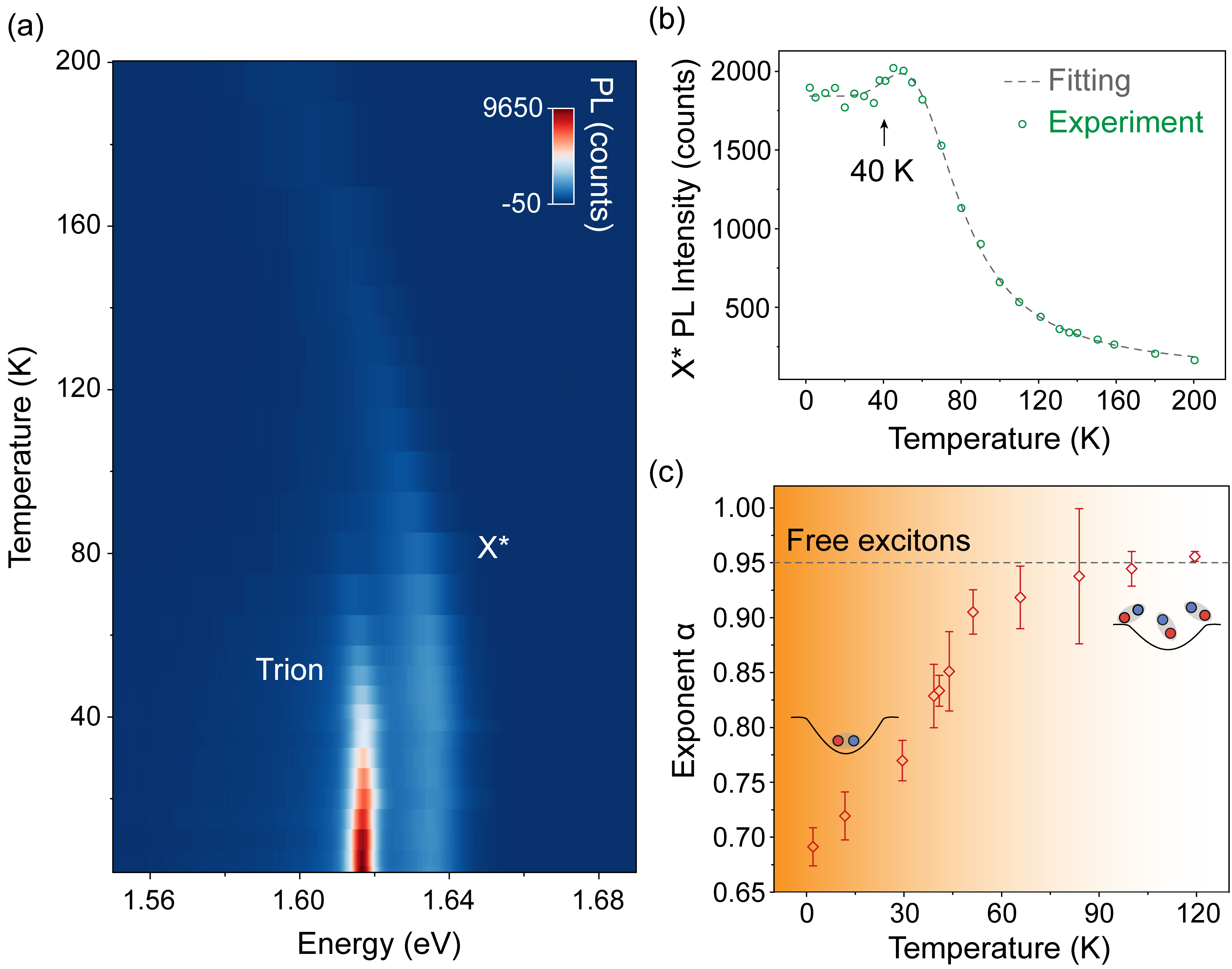}
	\caption{Temperature-dependent PL spectra. (a) PL spectra of device 1 at various temperatures. (b) Extracted PL intensity of X$^*$ peak as a function of temperature. The green open circles are the experimental data, and the gray dashed line represents the Arrhenius fitting. (c) Fitted power-law exponent $\alpha$ from the power-dependent measurement at different temperatures. The dashed line represents the $\alpha$ value of free excitons in monolayer MoSe$_2$ measured at 2 K. The left and right insets are schematics showing exciton confinement conditions relative to the CrSBr defect potential at low and high temperatures, respectively.}
	\label{Figure2}
\end{figure}

Temperature-dependent PL measurements are crucial for identifying the characteristics of localized excitons. Previous studies have indicated that as the temperature increases, localized excitons gradually escape the potential trap due to thermal fluctuations, leading to a much faster intensity decay compared to free excitons\cite{joshi2020localized,fang2023localization,zhang2017defect}. Interestingly, the intensity of the X$^*$ peak does not immediately quench with increasing temperature but persists up to temperatures exceeding 200 K, even dominating the spectra after weakening of the trion emission (Fig. \ref{Figure2}(a)). Additionally, the extracted PL intensity exhibits a non-monotonic behavior, showing a slight increase around 40 K, as shown in Fig. \ref{Figure2}(b). To gain further understanding of the temperature-dependent properties, we fit the peak intensity of X$^*$ using a modified Arrhenius formula\cite{joshi2020localized}:
\begin{equation}
    I(T)=I(0)\frac{1+Ae^{-E_{a1}/k_BT}}{1+Be^{-E_{a2}/k_BT}}
\end{equation}
where $I(0)$ is the PL intensity at $T=$ 0 K, A and B are fitting parameters describing the ratio of non-radiative decay rate to radiative decay rate, and $k_B$ is the Boltzmann constant. $E_{a1}$ represents the activation energy responsible for enhancing radiative recombination, while $E_{a2}$ is associated with the activation energy for thermal quenching processes at higher temperatures, mainly mediated by nonradiative pathways. For defect-trapped localized excitons, the parameter $E_{a2}$ is typically used to estimate the depth of the potential. However, our fitting result for $E_{a2}=35.8$ meV is significantly larger than the defect potentials reported previously ($<10$ meV) and strongly matched the thermal activation energy of neutral excitons in monolayer MoSe$_2$\cite{joshi2020localized}. Based on the observations, we propose two hypotheses: (1) The potential responsible for trapping excitons in MoSe$_2$ is remarkably deep ($\geq$ 35.8 meV), with thermal quenching limited solely by exciton binding energy. (2) The X$^*$ peak at higher temperatures actually originates from free exciton emissions, with energy closely matching that of localized excitons, making them indistinguishable in the spectra. We exclude the first hypothesis for two reasons. Firstly, it fails to explain the observed non-monotonic behavior of intensity with increasing temperature, as there is no discernible mechanism to increase the recombination rate of localized excitons. Secondly, the presence of such deep trapping potentials, as proposed, would give rise to a series of quantized energy levels with large energy separations, leading to the manifestation of multiple peaks from radiative recombination involving excitons occupying different energy levels\cite{tran2019evidence,fang2023localization}. However, from our PL spectra, the X$^*$ peak can be well fitted with a single Gaussian function at low temperatures (see Supplemental Material Fig. S5), showing no signs of multiple peaks. 

\begin{figure*}
	\centering
	\includegraphics[width=1\linewidth]{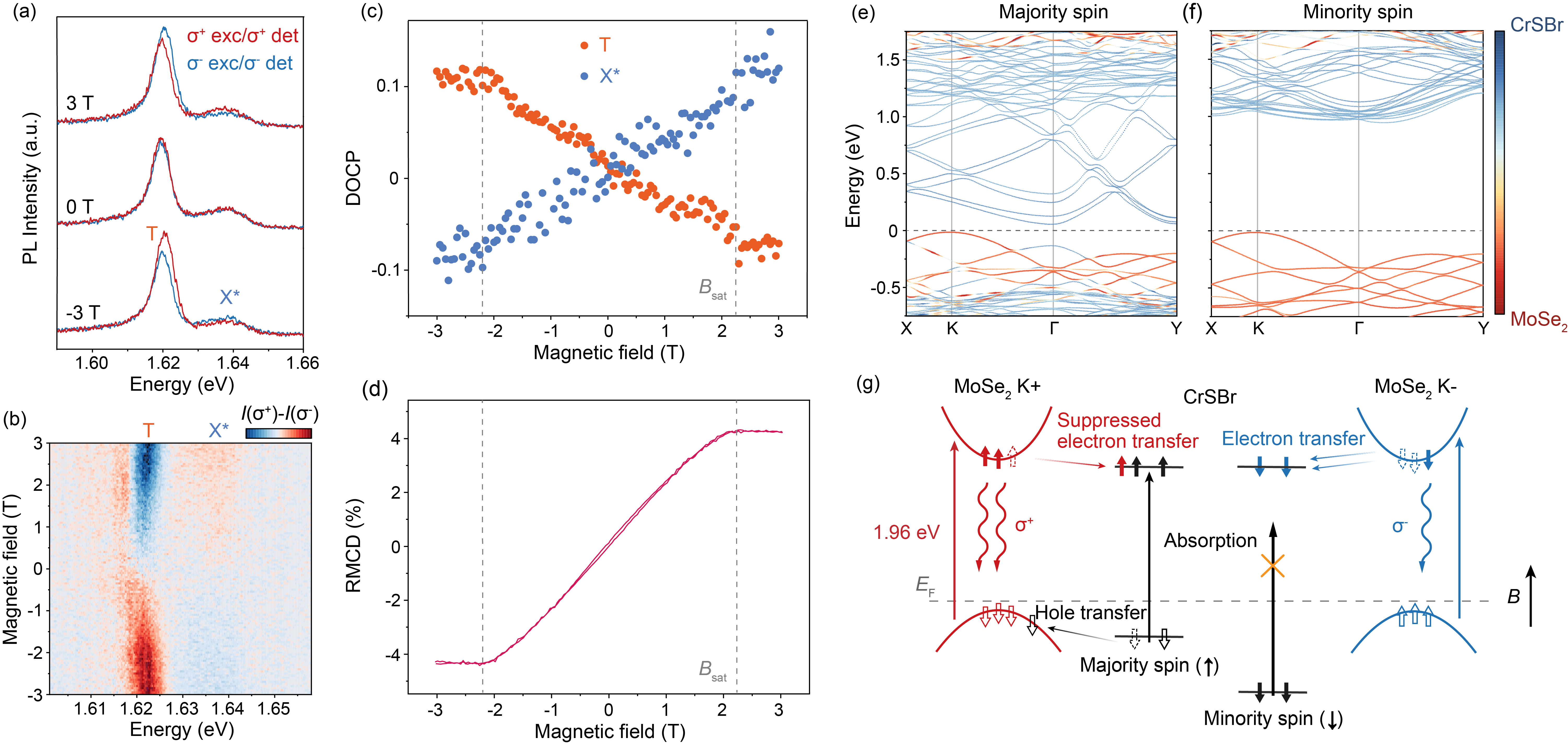}
	\caption{Charge transfer-mediated distinct valley polarization between X$^*$ and trions. (a) Right and left circularly polarized PL spectra at selective magnetic fields. (b) The difference in PL intensity between $\sigma^+$ and $\sigma^-$ polarizations in the energy and magnetic field parameter space. (c) DOCP as a function of magnetic field, with X$^*$ (trion) peak plotted as blue (orange) circles. The dashed lines represents the saturation field of CrSBr ($B_\text{sat}$). (d) RMCD signal of a CrSBr flake with similar thickness as a function of magnetic field along the \textit{c}-axis. (e), (f) DFT calculated band structure of the majority spin (e) and minority spin (f) of the  MoSe$_2$/FM CrSBr heterostructure. The blue and red bands represent CrSBr and MoSe$_2$ orbitals, respectively. The yellow bands indicate the orbitals where the two materials hybridize. (g) Schematic illustration of absorption, emission, and spin-dependent charge transfer processes in the heterostructure. }
	\label{Figure3}
\end{figure*}

Therefore, we conclude the second hypothesis, which suggests that as the temperature increases, the emission of X$^*$ gradually transitions from localized excitons to free excitons. The power dependence characteristics of the X$^*$ peak during the temperature evolution further support this scenario. From Fig. \ref{Figure2}(c), it can be observed that within the same excitation power range, the power law exponent $\alpha$ rapidly rises and exceeds $\alpha$ = 0.90 around 50 K, then gradually saturates and approaches 1. This indicates the presence of exciton delocalization processes at higher temperatures as well as a mixture of localized excitons and free excitons. Since $\alpha$ for free excitons measured in monolayer MoSe$_2$ at 2 K is approximately 0.95 (see Supplemental Material Fig. S4), we deduce that above 50 K, most excitons disengage from the confinement of defect potentials, and the emission of X$^*$ is mainly dominated by free excitons. The anomalous intensity rise of the X$^*$ peak with temperature can also be explained within this scenario, as the decrease in the population of localized excitons within defect potentials leads to an increase in the recombination of free excitons. 

After investigating the origin of X$^*$, we further explore its correlation with the magnetic properties of CrSBr. Fig. \ref{Figure3}(a) displays the circular polarization-resolved PL spectra of device 2 under selective magnetic fields perpendicular to the sample plane. Measurements were conducted in a co-polarized excitation/detection configuration, which could be either right ($\sigma^+$/$\sigma^+$) or left ($\sigma^-$/$\sigma^-$) circularly polarized. At a zero magnetic field, the $\sigma^+$ and $\sigma^-$ polarized PL spectra are nearly identical, as expected due to the in-plane magnetization of CrSBr and the preserved valley degeneracy of MoSe$_2$. Upon increasing the magnetic field to $\pm$ 3 T, clear polarization and energy splitting were observed in both the trion and X$^*$ peaks, indicating the breaking of valley degeneracy and time-reversal symmetry. Fig. \ref{Figure3}(b) depicts the difference in PL intensity between $\sigma^+$ and $\sigma^-$ polarized emission in the parameter space of the magnetic field and the emission energy. Notably, the intensity contrast of the trion and X$^*$ exhibits opposite dependences with the magnetic field. Valley polarization can be quantified by calculating the degree of circular polarization (DOCP), as DOCP $=(I_{\sigma^+}-I_{\sigma^-})/(I_{\sigma^+}+I_{\sigma^-})$, where $I_{\sigma^+} (I_{\sigma^-})$ represents the PL intensity with $\sigma^+$($\sigma^-$) polarization. The PL spectra at each magnetic field were fitted with two Gaussian functions to extract the DOCP of the trion and X$^*$ (see Supplemental Material Fig. S5). As depicted in Fig. \ref{Figure3}(c), both peaks exhibit a linear but opposite correlation of DOCP with the magnetic field, saturating around $\pm$ 2.2 T. Reflective magnetic circular dichroism (RMCD) spectroscopy (Fig. \ref{Figure3}(d)) confirms that the underlying CrSBr flake undergoes a spin-canting process under a perpendicular magnetic field and is fully spin-polarized at the saturation field $B_\text{sat}$ = $\pm$ 2.2 T\cite{wilson2021interlayer,telford2022coupling}. The magnetic field dependence of DOCP mirrors that of CrSBr RMCD signal, confirming the proximity interaction of the T and X$^*$ in MoSe$_2$ with the magnetic order of CrSBr.   

\begin{figure*}
	\centering
	\includegraphics[width=1\linewidth]{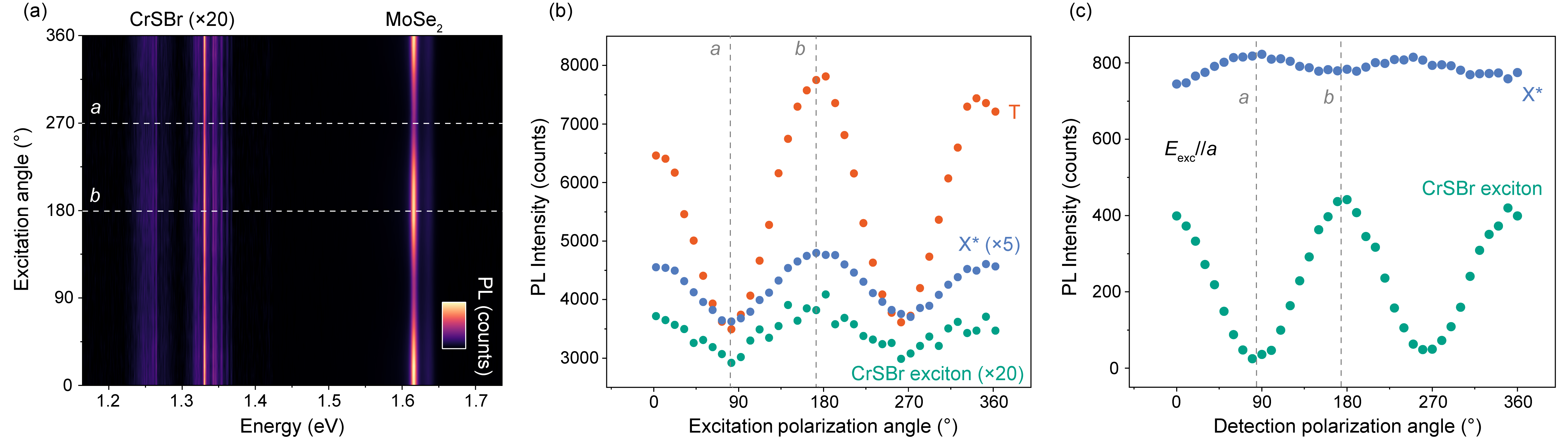}
	\caption{Proximity-induced anisotropic PL of MoSe$_2$/CrSBr heterostructure. (a) Excitation polarization angle-dependent PL spectra of the heterostructure. The white dashed lines represent the corresponding angles along the \textit{a}- and \textit{b}-axis, respectively. (b) Extracted PL intensity of trion (orange), X$^*$ (blue) and CrSBr exciton (green) as a function of excitation polarization angle. (c) Polarization-resolved PL intensity of X$^*$ and CrSBr exciton with excitation along the \textit{a}-axis. }
	\label{Figure4}
\end{figure*}

To gain a deeper understanding of the aforementioned phenomena, we investigate the band structure of the heterostructure. Density functional theory (DFT) calculations reveal a type-II band alignment between MoSe$_2$ and CrSBr (see Supplemental Material Fig. S7(a)), with the Fermi level situated close to the valence band maximum (VBM) of MoSe$_2$. Gate tunable reflection spectra confirmed the significant hole doping in the MoSe$_2$ monolayer (see Supplemental Material Fig. S6 and Fig. S7(b)). Fig. \ref{Figure3}(e) and \ref{Figure3}(f) illustrate the majority and minority spin band structures of MoSe$_2$/ferromagnetic (FM) CrSBr, corresponding to the spin-polarized states under an out-of-plane magnetic field. It is evident that the minority spin band of CrSBr exhibits a larger absorption band gap, exceeding the excitation energy of 1.96 eV (also see Supplemental Material Fig. S8). Consequently, as shown in Fig. \ref{Figure3}(g), under a positive magnetic field, more spin-up electrons are optically excited to the majority spin conduction band. The accumulation of electrons in the majority spin conduction band will impede the transfer of electrons from the MoSe$_2$ K+ valley, while the unoccupied states in the minority spin conduction band facilitate the transfer of spin-down electrons from the MoSe$_2$ K- valley, leading to a reduction in $\sigma^-$ polarized X$^*$ emission. Meanwhile, the photo-generated holes in the majority spin valence bands promote hole transfer to the K+ valley. Due to the inter-valley nature of MoSe$_2$ trions\cite{li2014valley}, the increase in holes within the K+ valley leads to a corresponding increase in $\sigma^-$ polarized trion emission. This elucidates the distinct and mirrored polarization behavior between X$^*$ and trion peaks.

It should be noted that, as stated in previous studies, an applied magnetic field can induce opposite valley polarization for excitons and trions in hole-doped MoSe$_2$\cite{li2014valley,lyons2020interplay}. However, in the MoSe$_2$/CrSBr heterostructure, the DOCP of trion and X$^*$ both exhibit discernible saturation trends above $B_\text{sat}$, implying the valley polarization is primarily influenced by the charge transfer processes driven by spin-polarized carriers. Nevertheless, owing to the in-plane easy-axis magnetization of the CrSBr layer, we observe negligible proximity-induced valley splitting, except for minor variations in the g-factor of trions around $B_\text{sat}$ (see Supplemental Material Fig. S9). This suggests a modest exchange splitting induced by the out-of-plane magnetization component of CrSBr. Similar results are also observed in device 1 (see Supplemental Material Fig. S10).

Inspired by the strong in-plane anisotropy of CrSBr, we further investigate the linear polarization properties of the MoSe$_2$/CrSBr heterostructure. Fig. \ref{Figure4}(a) presents the PL spectra of device 1 at varying excitation polarization angles. In stark contrast to the isotropic PL behavior of monolayer MoSe$_2$, the X$^*$ and trion peaks exhibit synchronized and prominent anisotropic responses with the emissions of CrSBr (Fig. \ref{Figure4}(b)). The highly anisotropic electronic structure of CrSBr results in a strong preference for absorption for light polarized along the \textit{b}-axis\cite{wilson2021interlayer,klein2023bulk}. Consequently, when excited with light polarized along the \textit{b}-axis, carrier accumulation in CrSBr hinders the electron transfer from MoSe$_2$ towards it (see Supplemental Material Fig. S11). As discussed earlier, the MoSe$_2$/CrSBr heterostructure manifests a type-II band alignment, where the reduction in charge transfer leads to an increase in the intensity of MoSe$_2$ X$^*$ and trion peaks under \textit{b}-polarized excitation. It is worth noting that the excitation linear dichroism (LD) ($I_\text{b}/I_\text{a}$) of CrSBr excitons is even smaller than that of MoSe$2$ emissions, with trion $\approx 2.14$, X$^* \approx 1.32$, and CrSBr exciton $\approx 1.22$. This counter-intuitive phenomenon may be attributed to the high excitation energy $E_{\text{exc}}$ = 1.96 eV, significantly exceeding the optical bandgap of CrSBr ($\sim$ 1.35 eV). Consequently, light polarized along the \textit{a}-axis partially undergoes absorption by higher energy states allowed optically in CrSBr, contributing to the PL emission after relaxation processes. Furthermore, the LD value measured for the bare CrSBr in the same device is 1.79 (see Supplemental Material Fig. S12), higher than that of the heterostructure, aligning with the physical picture of reduction in LD due to the charge transfer from MoSe$_2$ to CrSBr.

Lastly, we measured the linear polarization of PL emissions. Regardless of the excitation polarization along the \textit{a}-axis (Fig. \ref{Figure4}c) or \textit{b}-axis (see Supplemental Material Fig. S13), the emission of CrSBr excitons exhibits strong anisotropic behavior polarized along the \textit{b}-axis, consistent with previous reports\cite{wilson2021interlayer}. Interestingly, the X$^*$ emisson displays an opposite anisotropic behavior with small linear polarization along the \textit{a}-axis. This phenomenon can be ascribed to the reabsorption within the heterostructure. In the X$^*$ peak energy, CrSBr primarily absorbs the \textit{b}-polarized component of the initially isotropic emission from MoSe$_2$, resulting in the slight polarization along the \textit{a}-axis observed in the X$^*$ peak.

\section{Conclusion}

In summary, our study has unveiled a distinct spectral feature, labeled as X$^*$, within the MoSe$_2$/CrSBr vdW heterostructures. Through a meticulous analysis encompassing power- and temperature-dependent PL measurements, we ascribe the origin of X$^*$ to the recombination of neutral excitons confined by the intrinsic defect potentials present in the underlying CrSBr layer. Given the prevalence of structural defects in 2D materials, our findings suggest that localized exciton may be a common characteristic in various TMD heterostructures. Furthermore, we have identified an intriguing opposite circular-polarization behavior between X$^*$ and trions, attributed to hole doping induced by charge transfer in MoSe$_2$. By integrating RMCD measurements, we have elucidated the relationship between DOCP of MoSe$_2$ emissions and the magnetic order of CrSBr, stemming from spin-dependent charge transfer across the heterointerface. Moreover, the pronounced anisotropy in CrSBr has been shown to induce a distinctive polarization dependence in MoSe$_2$ emissions. Our work illuminates the pivotal role of defects in engineering optical properties and provides deeper insights into to the proximity effects involving different excitonic species in TMD heterostructures.
 
\section{Methods}
\noindent
\textbf{Crystal growth.} The bulk MoSe$_2$ single crystal was grown using the chemical vapor transport method. A mixture of Mo and Se powders with a stoichiometric ratio of 1:2 was heated to 700 $^{\circ}$C in 48 h and kept for 120 h to synthesize polycrystalline MoSe$_2$. The obtained precursor was carefully ground into powder, and then 1000 mg MoSe$_2$ powder was weighed and 14 mg/ml of Se was added as the self-transport agent. The thoroughly mixed powder was then sealed in a quartz tube under vacuum. Subsequently, the quartz tube was placed into a dual-zone furnace and grown for 216 hours under a temperature gradient from 1020 $^{\circ}$C to 960 $^{\circ}$C, with a heating/cooling rate of 1 $^{\circ}$C/min. To grow bulk CrSBr single crystal, a mixture of Cr (Alfa, 99.996\%) and TeBr$_4$ (Alfa, 99.999\%) with a molar ratio of 1:1.5 was sealed in a 12 cm long evacuated ($1.5\times10^{-4}$ Torr) quartz tuber. The tube was then subjected to a temperature gradient from 750 $^{\circ}$C to 450 $^{\circ}$C for 168 hours to obtain CrBr$_3$ crystals. Subsequently, a mixture containing Cr (87.0 mg, Alfa, 99.996\%), S (98.0 mg, Alfa, 99.99\%), and CrBr$_3$ (402.0 mg) was sealed in a new 12 cm long quartz tube. The quartz tube was placed in a dual-zone tube furnace. On the source side (sink side), the temperature was raised to 860 $^{\circ}$C (960 $^{\circ}$C) within 24 hours, maintained for 24 hours, then increased (decreased) to 960 $^{\circ}$C (860 $^{\circ}$C) within 12 hours and held for 48 hours before cooling to room temperature in 6 hours. 

\bigskip

\noindent
\textbf{Sample Fabrication.}
The MoSe$_2$ and CrSBr flakes were mechanically exfoliated onto SiO$_2$/Si substrates in ambient conditions and subsequently transferred into a glove box for further processing. Monolayer MoSe$_2$ was identified based on optical contrast and further verified through PL measurements conducted at room temperature. The MoSe$_2$/CrSBr heterostructures were meticulously assembled layer-by-layer transfer using a poly (bisphenol A carbonate) film atop a 
 polydimethylsiloxane (PDMS) stamp. Metallic Cr/Au electrodes were prepared through a series of steps involving electron-beam lithography, reactive ion etching (utilizing a plasma of the CHF$_3$/O$_2$ mixture), electron beam evaporation, and a lift-off process.
\bigskip

\noindent
\textbf{Optical Measurements.} The optical measurements were conducted within a closed-cycle helium cryostat (attoDRY2100) with a base temperature of 1.6 K and the capability to apply an out-of-plane magnetic field of up to 9 T. PL measurements utilized a HeNe laser (1.96 eV) focused onto the sample through a high numerical aperture (0.82) objective, resulting in a beam spot size of approximately 1 $\mu$m in diameter. The emitted light was collected by a spectrometer (SpectraPro HRS-500S), dispersed via a grating with a groove density of 600 mm$^{-1}$, and detected using a liquid-nitrogen-cooled charge-coupled device (PyLoN:400). Spatial mapping of the PL signal was achieved by manipulating the piezo sample stage. In the RMCD measurements, the incident linearly polarized light from the HeNe laser was modulated between left and right circular polarization by a photoelastic modulator (PEM) before being directed onto the sample at normal incidence. The reflected light was captured by a photomultiplier tube (THORLABS PMT1001/M). The RMCD signal was determined by the ratio of the a.c. component of PEM at 50.052 kHz to that of a chopper at around 756 Hz (detected using a two-channel lock-in amplifier Zurich HF2LI). For reflection spectrum measurements, a halogen lamp served as the white light source, with its output focused through a 50 $\mu$m diameter pinhole (JCOPTIX MPS1-50) and then collimated by an objective lens. The resulting beam spot diameter was approximatedly 2 $\mu$m, with an excitation power around 100 nW. 
\bigskip

\noindent
\textbf{DFT Calculations.} DFT calculations were conducted using VASP. The calculations employed a basis set of projector augmented plane waves with a plane wave energy cutoff set at 400 eV. The PBE functional was chosen to describe the exchange-correlation interaction, with spin-orbit coupling (SOC) disabled for this study. To ensure compatibility between MoSe$_2$ and CrSBr lattices, the lattice of MoSe$_2$ was stretched by 2\%. A $k$-mesh of $3\times3\times1$ was utilized to sample the momentum space. Additionally, a vacuum space exceeding 15 \r{A} was incorporated in directions perpendicular to the material plane to prevent interactions from periodic images. 
	
\begin{acknowledgments}
This work was supported by the National Key R\&D Program of China (No. 2022YFA1203902), the National Natural Science Foundation of China (No. 12250007), Beijing Natural Science Foundation (No. JQ21018), and the China Postdoctoral Science Foundation (2023TQ0003 and 2023M740122). K.W. and T. T. acknowledge support from the JSPS KAKENHI (Grant Number 21H05233 and 23H02052) and World Premier International Research Center Initiative (WPI), MEXT, Japan.

\end{acknowledgments}
	
\bibliography{Reference}
	
\setcounter{figure}{0}
\setcounter{table}{0}
\newpage

	\nocite{*}
	
\end{document}


\date{}
\onecolumn{
\maketitle 
\vspace{-5mm}
\begin{center}
\begin{minipage}{1\textwidth}
\begin{center}

\textit{ \small{
\textsuperscript{1} State Key Laboratory for Artificial Microstructure $\rm{\&}$ Mesoscopic Physics and Frontiers Science Center for Nano-Optoelectronics, School of Physics, Peking University, Beijing 100871, China
\\\textsuperscript{2} Academy for Advanced Interdisciplinary Studies, Peking University, Beijing 100871, China
\\\textsuperscript{3} John A. Paulson School of Engineering and Applied Sciences, Harvard University, Cambridge, Massachusetts 02138, United States
\\\textsuperscript{4} Research Center for Electronic and Optical Materials, National Institute for Materials Science, 1-1 Namiki, Tsukuba 305-0044, Japan
\\\textsuperscript{5} Research Center for Materials Nanoarchitectonics, National Institute for Materials Science, 1-1 Namiki, Tsukuba 305-0044, Japan
\\\textsuperscript{6} School of Semiconductor Science and Technology, South China Normal University, Foshan 528225, China
\\\textsuperscript{7} Yangtze Delta Institute of Optoelectronics, Peking University, Nantong 226010 Jiangsu, China
\\\textsuperscript{8} Liaoning Academy of Materials, Shenyang, 110167, China
\\{$\dagger$} Emails: yang\_shiqi@pku.edu.cn, chenzuxin@m.scnu.edu.cn, ye\_yu@pku.edu.cn\\}
\vspace{5mm}
}

\end{center}
\end{minipage}
\end{center}

\noindent 
\textbf{\large{Contents:}}
\bigskip

\noindent
I. Characterization of CrSBr emissions in device 1.
\bigskip

\noindent
II. Spectral characteristics of different MoSe$_2$/CrSBr devices.
\bigskip

\noindent
III. Spatial PL mappings of device 1.
\bigskip

\noindent
IV. Power dependence of neutral exciton in bare MoSe$_2$.
\bigskip

\noindent
V. Details of the PL fitting.
\bigskip

\noindent
VI. Reflection spectra of monolayer MoSe$_2$ and MoSe$_2$/CrSBr heterostructure.
\bigskip

\noindent
VII. Theoretical and experimental investigations of band alignment.
\bigskip

\noindent
VIII. Valley splittings of trion and X$^*$ peaks of device 2.
\bigskip

\noindent
IX. Excitonic valley polarizations and valley splittings of device 1.
\bigskip

\noindent
X. Schematic illustrating the excitation linear dichroism of MoSe$_2$/CrSBr heterostructure.
\bigskip

\noindent
XI. Comparison of excitation linear dichroism between bare CrSBr and heterostructure.
\bigskip

\noindent
XII. Polarization of X$^*$ and CrSBr exciton with excitation polarization along the \textit{b}-axis.

\clearpage
\newpage 

\setlength\parindent{12pt}

\noindent
\textbf{I. Characterization of CrSBr emissions in device 1.}
\bigskip

We also investigated the spectral characteristics of CrSBr. Fig.\ref{crsbr}(a) presents the photoluminescence (PL) spectra of the heterostructure and bare CrSBr in the CrSBr emission window. Unlike the case of MoSe$_2$ emissions, the intensity and energy of the peaks remain almost unchanged, except for a small linewidth broadening observed in the heterostructure due to increased scattering or charge transfer processes. Moreover, these emission peaks exhibit no observable changes with the applied gate voltage (Fig.\ref{crsbr}(b)).

\begin{figure*}[ht!]
	\centering
	\includegraphics[width=1\textwidth]{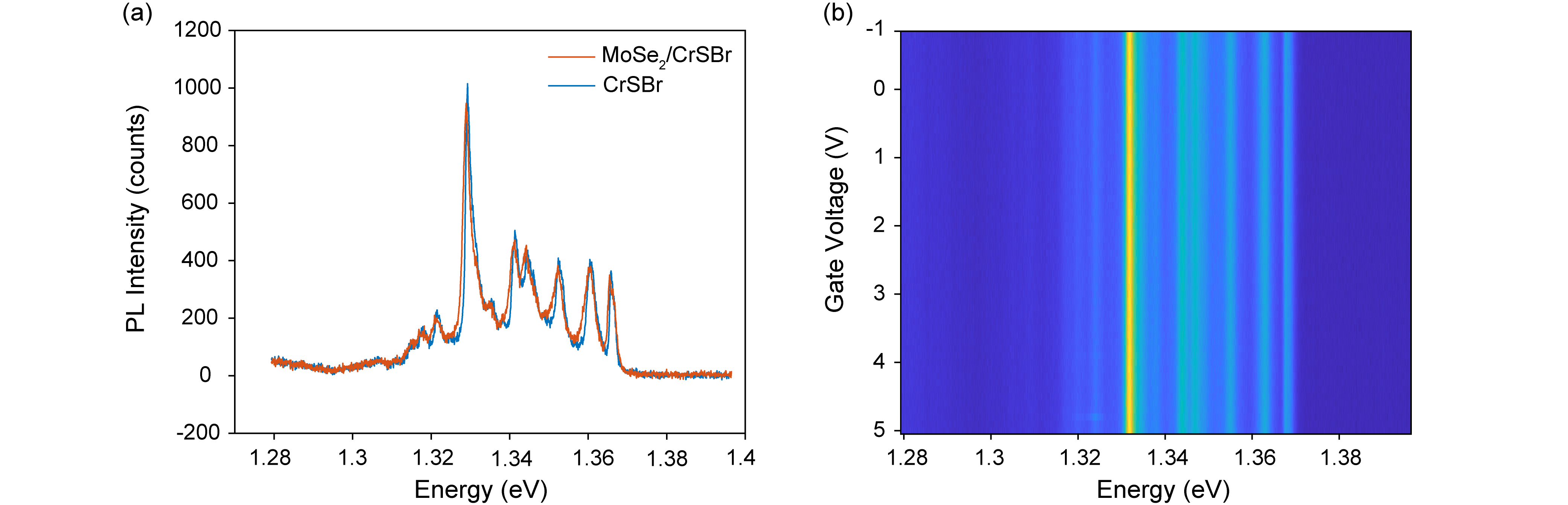}
	\caption{\textbf{PL spectra of CrSBr emissions in device 1.} 
        \textbf{(a),} CrSBr emissions from the heterostructure (orange) and bare CrSBr (blue), respectively. 
        \textbf{(b),} Gate voltage-dependent CrSBr PL spectrum from the heterostructure.
	}
	
	\label{crsbr}
\end{figure*}

\newpage
\noindent 
\textbf{II. Spectral characteristics of different MoSe$_2$/CrSBr devices.}
\bigskip

\begin{figure*}[ht!]
\centering
\includegraphics[width=1\textwidth]{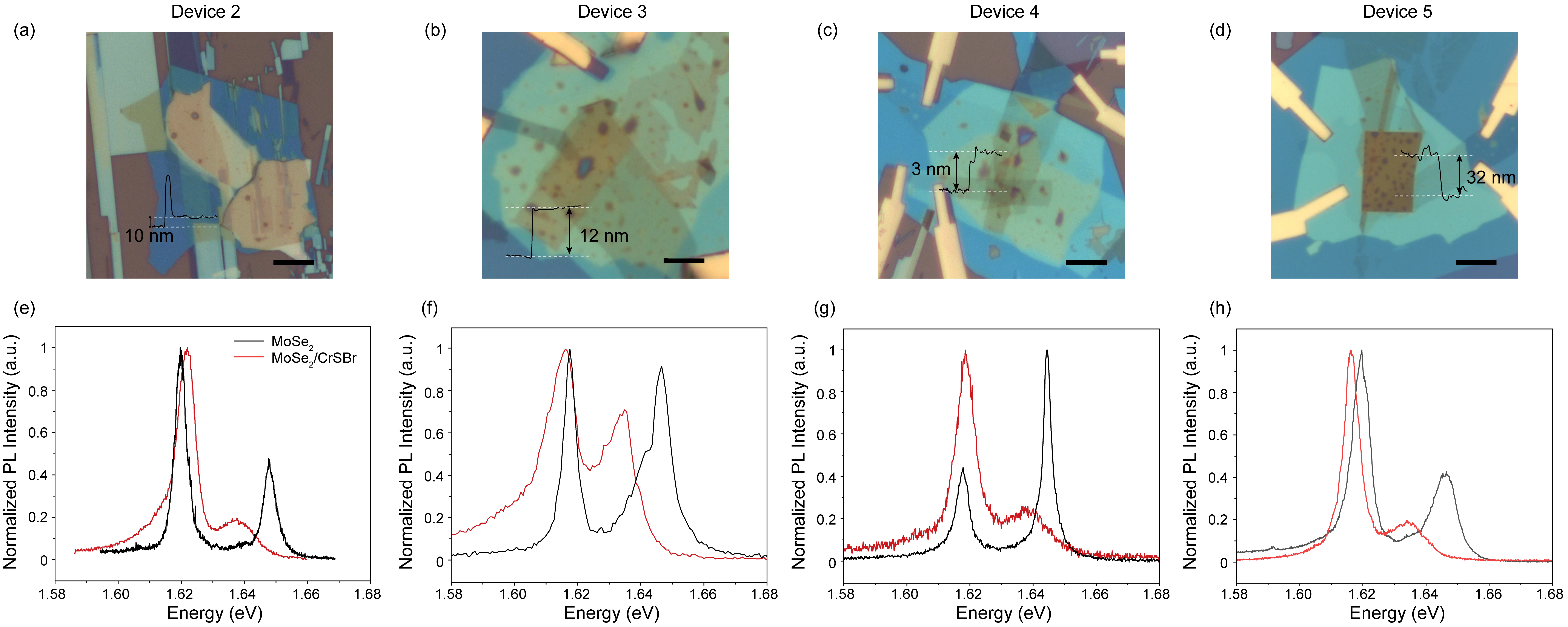}
\caption{\textbf{Basic characterizations and PL spectra of device 2 to device 5.} 
\textbf{(a)-(d),} Optical images of device 2 to device 5. The inset line profile presents the thickness of CrSBr in the heterostructure measured by atomic force microscopy. All scale bars: 10 $\mu$m.
\textbf{(e)-(h),} PL spectra of device 2 to device 5 with excitation spot on the bare MoSe$_2$ (black) and heterostructure (red) regions, respectively. Similar to the results of device 1 presented in the main text, a new peak appeared with its energy lies between the trion peak and the neutral exciton peak of the pristine MoSe$_2$. 
}

\label{devices}
\end{figure*}

\begin{table}[ht!]
\centering
\caption{List of localized exciton energy (X$^*$), trion energy (T) of the MoSe$_2$/CrSBr heterostructure, and neutral exciton energy (X$^0$), trion energy (T$^0$) of bare MoSe$_2$, and the corresponding energy difference between X$^*$ and X$^0$, T and T$^0$ of the five measured devices. The disparities in the absolute emission energy observed among different devices may stem from fluctuations in the dielectric environment and doping level.}

\renewcommand\arraystretch{2}
\begin{tabular}{lcccccc}
	\hline
        & X$^*$ (eV) & T (eV)  & X$^0$ (eV) & T$^0$ (eV) & X$^*$$-$X$^0$ (meV) & T$-$T$^0$ (meV)  \\ \hline
	Device 1   & 1.635  & 1.617 & 1.648  & 1.618  & $-$13 & $-$1 \\
	Device 2   & 1.638  & 1.622 & 1.648  & 1.62   & $-$10 & 2 \\
	Device 3   & 1.635  & 1.615 & 1.647 & 1.617   & $-$12  & $-$2  \\ 
        Device 4   & 1.637  & 1.618 & 1.645 & 1.618   & $-$8  & $\sim 0$  \\
        Device 5   & 1.634  & 1.616 & 1.646 & 1.619   & $-$12  & $-$3  \\
 \hline
\end{tabular}
\end{table}

\newpage
\noindent
\textbf{III. Spatial PL mappings of device 1.}
\bigskip

We characterized the spatial variations of the PL intensity in device 1. The integrated intensity maps of the X$^*$ peak and the trion peak are presented in Figs. \ref{PLmap}(c) and (d), respectively. As depicted in Fig. \ref{PLmap}(c), the X$^*$ emission dominates the heterostructure region and exhibits strong spatial uniformity. Meanwhile, significant quenching of trion emission occurs in the heterostructure region compared to bare MoSe$_2$, indicating an efficient interlayer charge transfer process.

\begin{figure*}[ht!]
	\centering
	\includegraphics[width=1\textwidth]{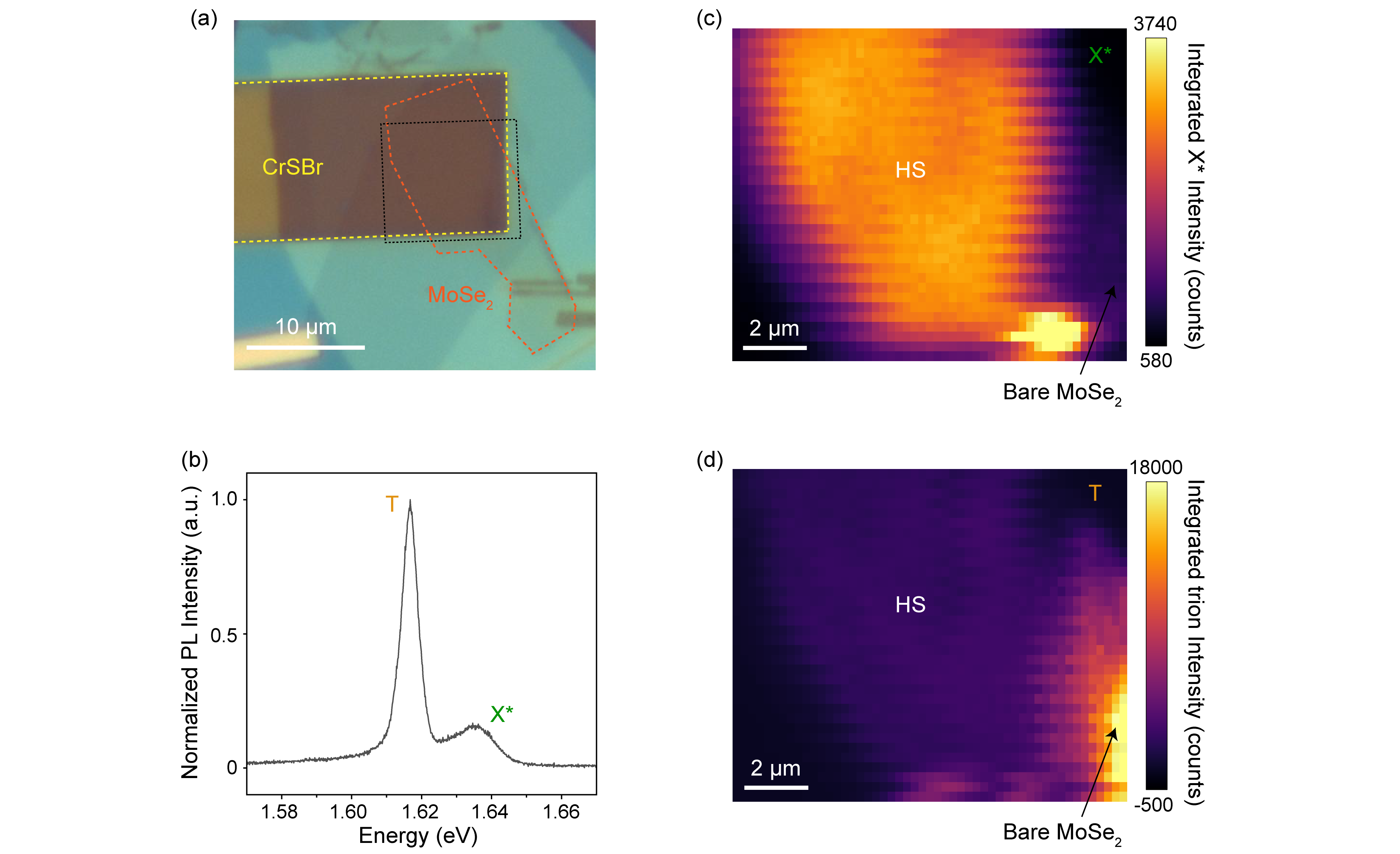}
	\caption{\textbf{Spatial mappings.} 
	\textbf{(a),} Optical image of device 1. The MoSe$_2$ monolayer and CrSBr flake are outlined in orange and yellow, respectively. The selected area for PL mapping is outlined by the dashed black box.
        \textbf{(b),} PL spectra of device 1. The trion peak and the X$^*$ peak are labeled.
        \textbf{(c), (d),} Spatial distributions of the integrated PL intensity of the X$^*$ (c) and the trion (d) peaks. For the X$^*$ peak, the selected energy range for integration is about 7 meV centered at the X$^*$ peak energy, and for the trion peak, the energy range is about 3 meV. 
  }
	\label{PLmap}
\end{figure*}

\newpage
\noindent
\textbf{IV. Power dependence of neutral exciton in bare MoSe$_2$.}
\bigskip

\begin{figure*}[ht!]
	\centering
	\includegraphics[width=1\textwidth]{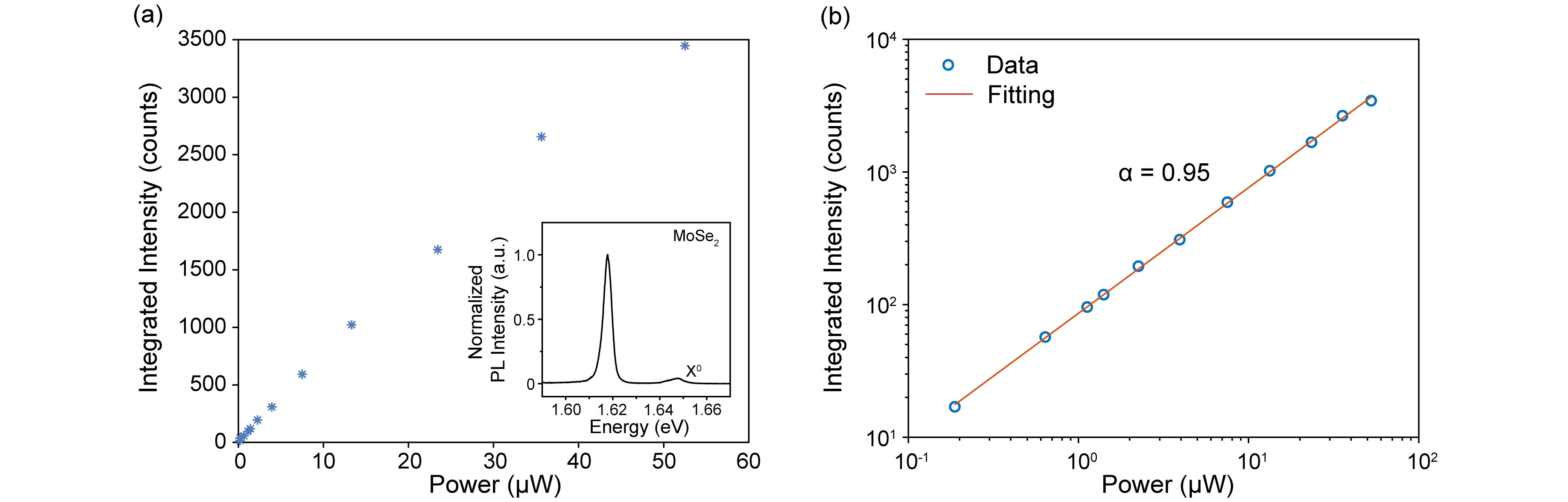}
	\caption{\textbf{Power-dependent PL intensity of MoSe$_2$ neutral exciton.} 
	\textbf{(a),} The integrated intensity of MoSe$_2$ neutral exciton as a function of excitation power. The selected energy range for integration is from 1.647 eV to 1.651 eV. The inset shows the PL spectra of the monolayer MoSe$_2$, with the neutral exciton peak labeled as X$^0$. 
        \textbf{(b),} Integrated exciton intensity \textit{versus} excitation power in the logarithmic coordinates. The blue open circles are the experimental data and the red line shows the linear fitting, the slope of which represents the power-law exponent $\alpha$ = 0.95.
	}
	
	\label{Calibration_1}
\end{figure*}

\bigskip
\noindent
\textbf{V. Details of the PL fitting.}

\bigskip

In the main text Fig. 1 and Fig. 3, spectral fitting extracts the peak intensity and energy to obtain the power- and magnetic field-dependent characteristics. Fig. \ref{specfit} shows the representative PL spectra of two MoSe$_2$/CrSBr devices (device 1 and device 2) with similar spectral features. Both spectra have two peaks, a trion and a localized exciton peak X$^*$. For device 1, we fit the spectra using a Gaussian profile for the X$^*$ peak and a pseudo-Voigt profile for the trion peak. For device 2, two Gaussian profiles are used to fit the X$^*$ and trion peaks. In the fitting process, we set a lower energy bound ( $\sim$ 1.61 eV) to avoid the complexities associated with the low-energy PL features\cite{zhong2020layer}.

\begin{figure*}[ht!]
	\centering
	\includegraphics[width=1\textwidth]{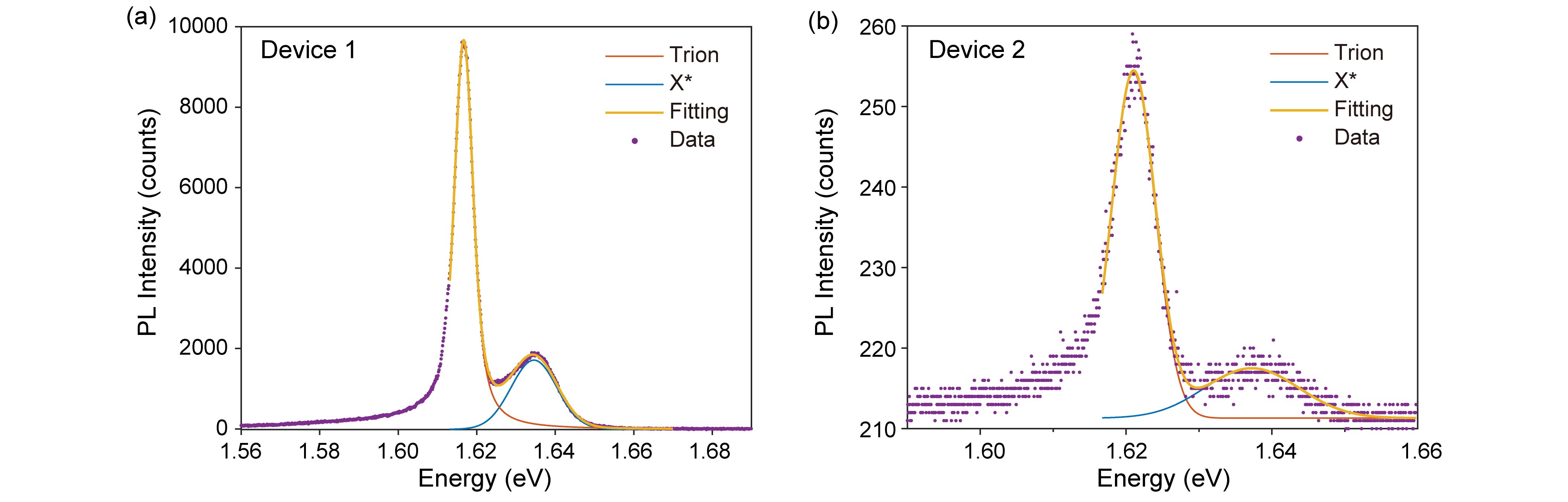}
	\caption{\textbf{PL spectra and their corresponding fittings.} 
    \textbf{(a),(b),} PL spectra and their corresponding fitting results of MoSe$_2$/CrSBr device 1 (a) and device 2 (b). The purple dots are the experimental data. The blue and orange lines are the fitted lineshapes of the X$^*$ peak and trion peak, respectively. The yellow line represents the total fit. 
		}
	\label{specfit}
\end{figure*}

\newpage
\noindent
\textbf{VI. Reflection spectra of monolayer MoSe$_2$ and MoSe$_2$/CrSBr heterostructure. }
\bigskip

Fig. \ref{Reflection}(a), (b) present the reflection spectra of bare MoSe$_2$ and MoSe$_2$/CrSBr, respectively. The multi-layer structure of our device leads to interference between light reflected at different interfaces, which introduces a background signal to the reflection spectrum. Because the phase of the reflection is wavelength-dependent, the background signal will vary with photon energy. Additionally, the interference between the excitonic resonance and the background signal significantly modifies the spectral lineshape. To address this effect, we first obtain the background signal by polynomial fitting at spectral regions away from the resonance. After subtracting the background, we fit the spectrum using an effective dispersive Lorentzian function\cite{smolenski2021signatures}:
\begin{equation}
    \frac{\Delta R}{R}(E)=Acos(\phi)\frac{\gamma /2}{(E-E_0)^2+\gamma ^2/4}+Asin(\phi)\frac{E_0-E}{(E-E_0)^2+\gamma ^2/4}+C
\end{equation}
where $E$ is the photon energy, $A$, $\gamma$, $E_0$, and C correspond to the amplitude, resonance energy, linewidth, and constant background, respectively. $\phi$ denotes the phase difference between the reflections. From the fitting, we obtain the excitonic resonance energy $E$=1.652 eV for monolayer MoSe$_2$, and $E$=1.643 eV for heterostructure. The $\sim$ 9 meV redshift of the heterostructure resonance compared to the bare MoSe$_2$ may result from the band gap reduction due to the interlayer couplings. In addition to the excitonic redshift, the resonance in heterostructure has a much broader linewidth due to the charge transfer between these two materials. We also observed a small feature at around 1.622 eV in the bare MoSe$_2$ reflection spectrum, which is assigned to the attractive polaron which has been extensively studied in the heavily doped MoSe$_2$ monolayer.

\begin{figure*}[ht!]
	\centering
	\includegraphics[width=1\textwidth]{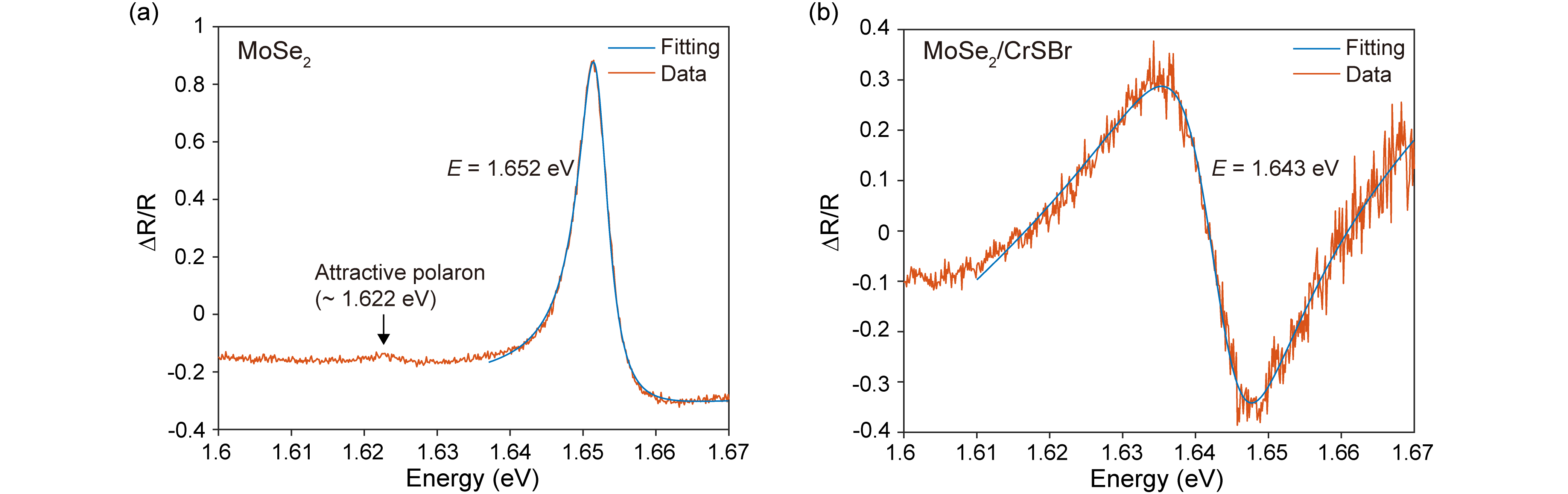}
	\caption{\textbf{Reflection spectra and fittings of device 1.} 
		\textbf{(a),(b),} Reflection spectra of monolayer MoSe$_2$ (a) and MoSe$_2$/CrSBr heterostructure (b), with the background signal already been subtracted. The solid orange lines are the experimental data and the blue lines represent the spectra fitting described in the text. The low-energy feature at around 1.622 eV in (a) is assigned to the attractive polaron.
	}
	
	\label{Reflection}
\end{figure*}

\newpage
\noindent
\textbf{VII. Theoretical and experimental investigations of band alignment.}
\bigskip

Fig. \ref{band1}(a) shows the DFT calculated band structure of MoSe$_2$/CrSBr heterostructure. The global conduction band minimum (CBM) and valence band maximum (VBM) originate from CrSBr and MoSe$_2$ respectively, which reveals a type-II band alignment. The Fermi level lies almost at the MoSe$_2$ VBM, suggesting a heavily hole-doped system. This is further verified by our gate-dependent reflection spectrum (Fig. \ref{band1}(b)). Unlike pristine MoSe$_2$\cite{liu2021exciton}, the lower-lying feature (only occurs at the negative gate voltage regime) observed in Fig. \ref{band1}(b) is attributed to the attractive polaron. As illustrated in Fig. \ref{band1}(c), the particular band alignment between MoSe$_2$ and CrSBr favors electron occupations in the CrSBr conduction bands at positive gate voltages, leading to a predominately hole-doped MoSe$_2$. Therefore, the attractive polaron at the electron-doped side is hard to observe.

\begin{figure*}[ht!]
	\centering
	\includegraphics[width=1\textwidth]{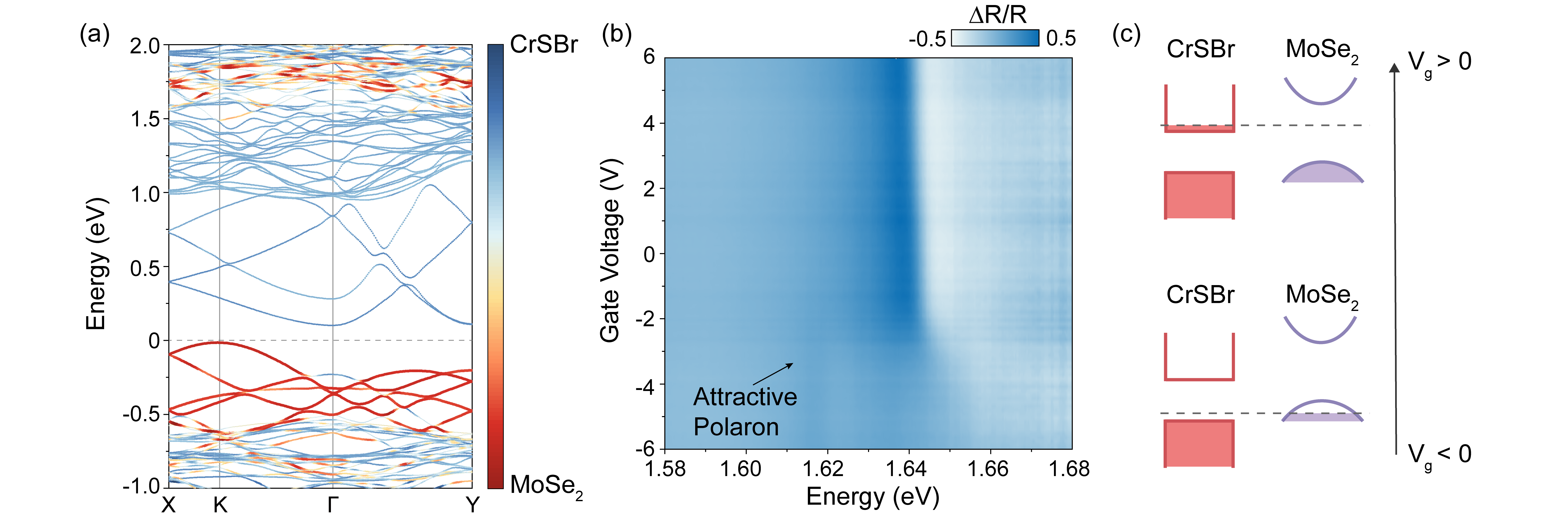}
	\caption{\textbf{Band structure and gate-dependent reflection spectrum of MoSe$_2$/CrSBr.} \textbf{(a),} DFT calculated band structure of the MoSe$_2$/antiferromagnetic (AFM) state CrSBr. The blue and red bands represent CrSBr and MoSe$_2$ orbitals, respectively. The yellow bands indicate hybridization between the two materials. \textbf{(b),} Gate voltage-dependent reflection spectrum of MoSe$_2$/CrSBr heterostructure. The lower-energy spectral feature is assigned to the attractive polaron. \textbf{(c),} Schematics of band alignment and doping levels of the heterostructure at positive and negative gate voltages.
        }
	
	\label{band1}
\end{figure*}

\newpage
Fig. \ref{DFT} shows the spin-dependent electronic band structure of MoSe$_2$/FM CrSBr in a larger energy range (from $-$2 eV to 2 eV) compared to Fig. 3(e),(f) that presented in the main text. The energy gap of the minority spin bands is much larger than that of the majority spin bands. The hybridized valence band orbitals (yellow bands) only appear at energy lower than $-$1 eV. Considering that DFT calculations usually underestimate the energy gap, we deduce that the absorption edge of the minority spin bands exceeds the excitation energy of 1.96 eV.

\begin{figure*}[ht!]
	\centering
	\includegraphics[width=1\textwidth]{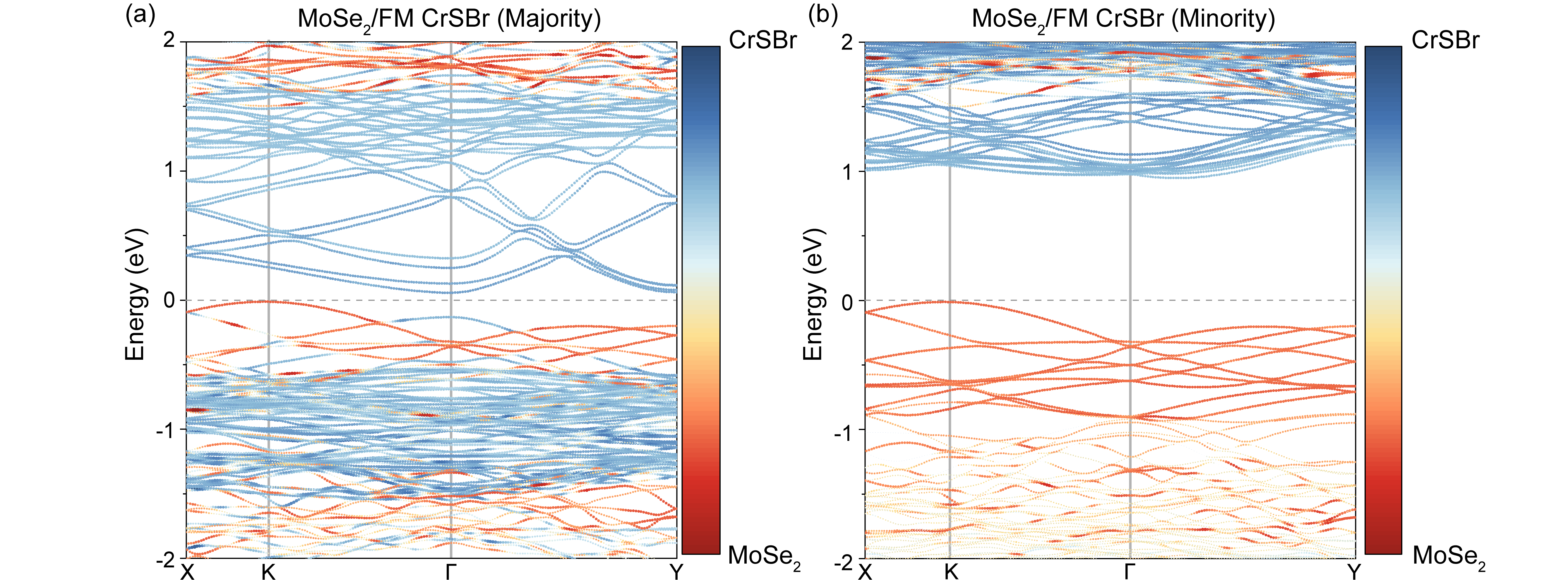}
	\caption{\textbf{Electronic band structure of MoSe$_2$/FM CrSBr in a larger energy range.} 
		\textbf{(a),(b),} DFT calculated the majority (a) and minority (b) spin band structures of MoSe$_2$/FM CrSBr. The energy range is from $-$2 eV to 2 eV. The horizontal grey dashed lines in both figures represent the Fermi level.
	}
	
	\label{DFT}
\end{figure*}

\newpage
\noindent
\textbf{VIII. Valley splittings of trion and X$^*$ peaks of device 2.}
\bigskip

We extract the energies of trion and X$^*$ peaks from the PL spectrum using the fitting process described in Supplementary Material IV. The valley splitting is defined as $\Delta = E_{\sigma^+}-E_{\sigma^-}$, where $E_{\sigma^+}$ ($E_{\sigma^-}$) is the $\sigma^+$/$\sigma^+$ ($\sigma^-$/$\sigma^-$) polarized PL peak energy. Fig. \ref{split} (a) and (b) present the valley splitting of the trion and the X$^*$ peaks, respectively. Interestingly, the slope of the trion valley splitting exhibits a clear reduction across the saturation field ($B_\text{sat}$). When $|B|>B_\text{sat}$, the CrSBr flake is fully spin-polarized and the valley splitting is expected to be dominated by the Zeeman effect. The fitted slope (0.11 meV/T when $B<-B_\text{sat}$, and 0.12 meV/T when $B>B_\text{sat}$) is consistent with the previously reported Zeeman slope of a monolayer MoSe$_2$ trion\cite{li2014valley}. By contrast, the slope of the valley splitting is much larger (about 0.28 meV/T) when $|B|<B_\text{sat}$ that corresponds to the CrSBr spin-canted state. This indicates that the out-of-plane component of the CrSBr magnetization introduces an additional proximity exchange field that contributes to the excitonic valley Zeeman splitting. For the X$^*$ peak, the valley splitting shows an overall linear dependence on the applied magnetic field. Unfortunately, we could not extract detailed behaviors from the data due to the low signal-to-noise ratio of the X$^*$ emission (see Fig. \ref{specfit}(b)).

\begin{figure*}[ht!]
	\centering
	\includegraphics[width=1\textwidth]{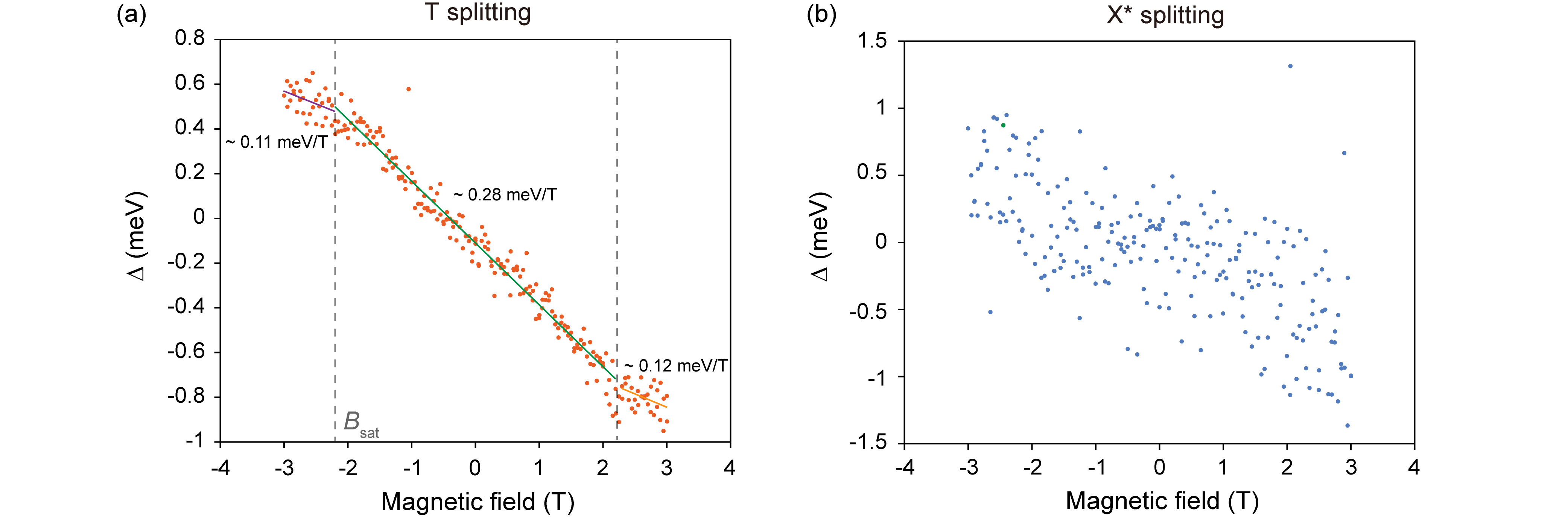}
	\caption{\textbf{Valley splitting of trion and X$^*$ peaks in device 2.} \textbf{(a),(b),} Energy splitting between $\sigma^+$ and $\sigma^-$ polarized trions (a) and X$^*$ (b). The grey dashed lines in (a) indicate the saturation field ($B_\text{sat}$) of CrSBr. The solid lines are the linear fits of the data.
    }
	
	\label{split}
\end{figure*}

\newpage
\noindent
\textbf{IX. Excitonic valley polarizations and valley splittings of device 1.}
\bigskip

The valley polarization and valley splitting of the trion and X$^*$ peaks of device 1 exhibit similar behaviors as observed in device 2. The degree of circular polarization (DOCP) of both peaks shows a magnetic field dependence analog to the CrSBr spin-canting process. Again, there exists an inversion in the polarity of the DOCP between the X$^*$ peak and the trion peaks. Furthermore, the valley splitting of both peaks undergoes a transition across $B\approx2$ T. The antisymmetric response and the oscillating signatures observed in the device could potentially arise from the sample drift by the presence of an external magnetic field.

\begin{figure*}[ht!]
	\centering
	\includegraphics[width=1\textwidth]{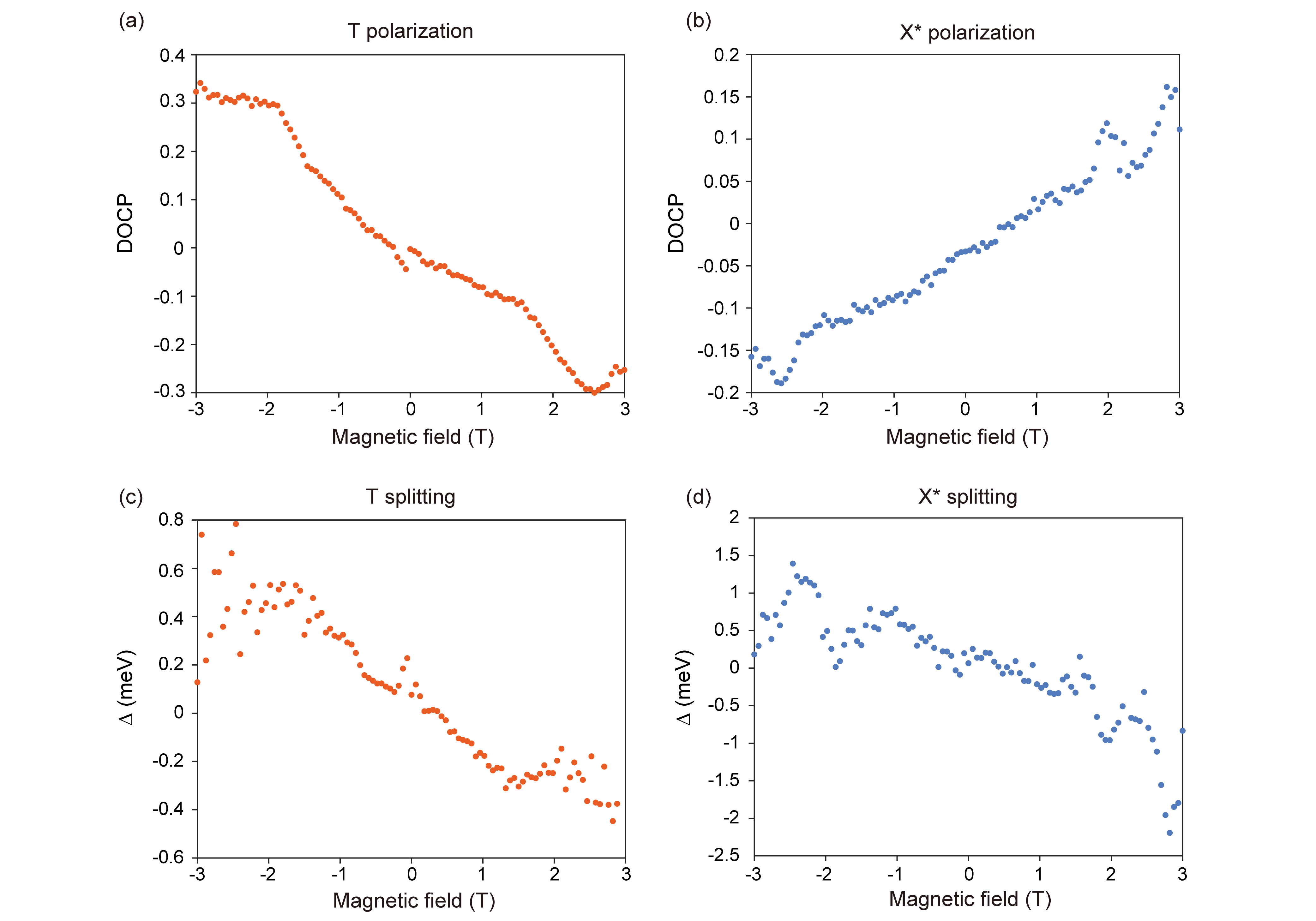}
	\caption{\textbf{Valley polarization and valley splitting of trion and X$^*$ peaks in device 1.} 
		\textbf{(a),(b),} Degree of circular polarization of trions (a) and X$^*$ (b). \textbf{(c),(d),} Valley splitting of trions (c) and X$^*$ (d).
	}
	
	\label{d1_pol}
\end{figure*}

\newpage
\noindent
\textbf{X. Schematic illustrating the excitation linear dichroism of MoSe$_2$/CrSBr heterostructure. }
\bigskip

\begin{figure*}[ht!]
	\centering
	\includegraphics[width=1\textwidth]{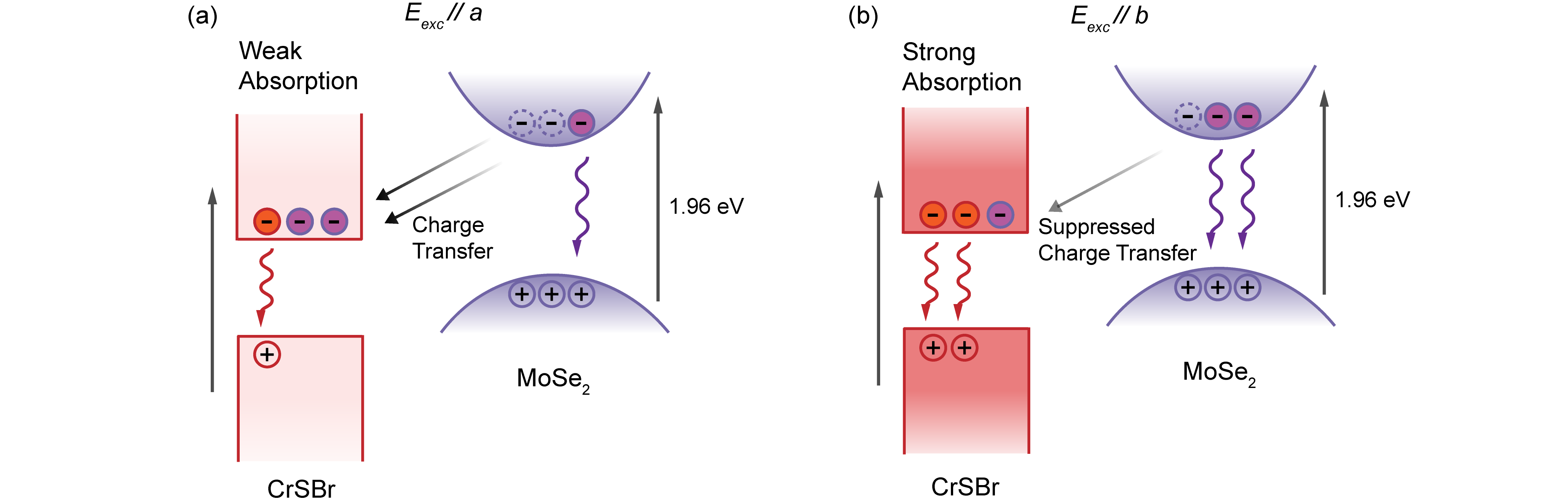}
	\caption{\textbf{Schematic illustration of excitation polarization-dependent emissions of the heterostructure.} 
		\textbf{(a),(b),} Schematic depicting the charge transfer and emission processes under excitation along the \textit{a}- (a) and \textit{b}-axis (b).
	}
	
	\label{schematic}
\end{figure*}

\noindent
\textbf{XI. Comparison of excitation linear dichroism between bare CrSBr and heterostructure.}

\bigskip
We measured the excitation linear dichroism (LD) ($I_\text{b}/I_\text{a}$) of CrSBr exciton (1.33 eV) on bare CrSBr and heterostructure. The LD value shows a clear reduction for MoSe$_2$/CrSBr (1.22) compared to that of bare CrSBr (1.79). This further verifies the existence of charge transfer and re-absorption processes in the heterostructure. 

\begin{figure*}[ht!]
	\centering
	\includegraphics[width=1\textwidth]{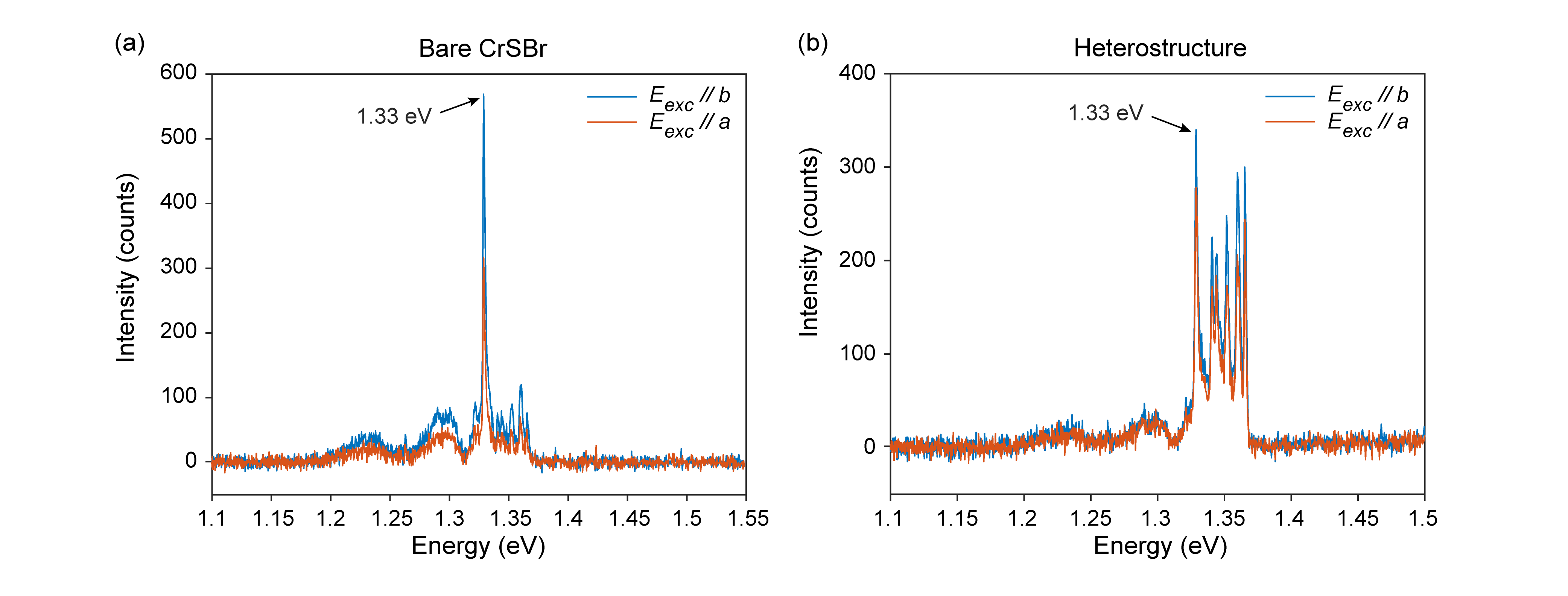}
	\caption{\textbf{Excitation polarization dependent-PL spectrum.} 
		\textbf{(a),(b),} PL spectra of bare CrSBr (a) and the heterostructure (b), with the excitation polarization along the \textit{a}- (orange) and \textit{b}-axis (blue).
	}
	
	\label{anisotropy}
\end{figure*}

\newpage
\noindent
\textbf{XII. Polarization of X$^*$ and CrSBr exciton with excitation polarization along the \textit{b}-axis.}
\bigskip

\begin{figure*}[ht!]
	\centering
	\includegraphics[width=1\textwidth]{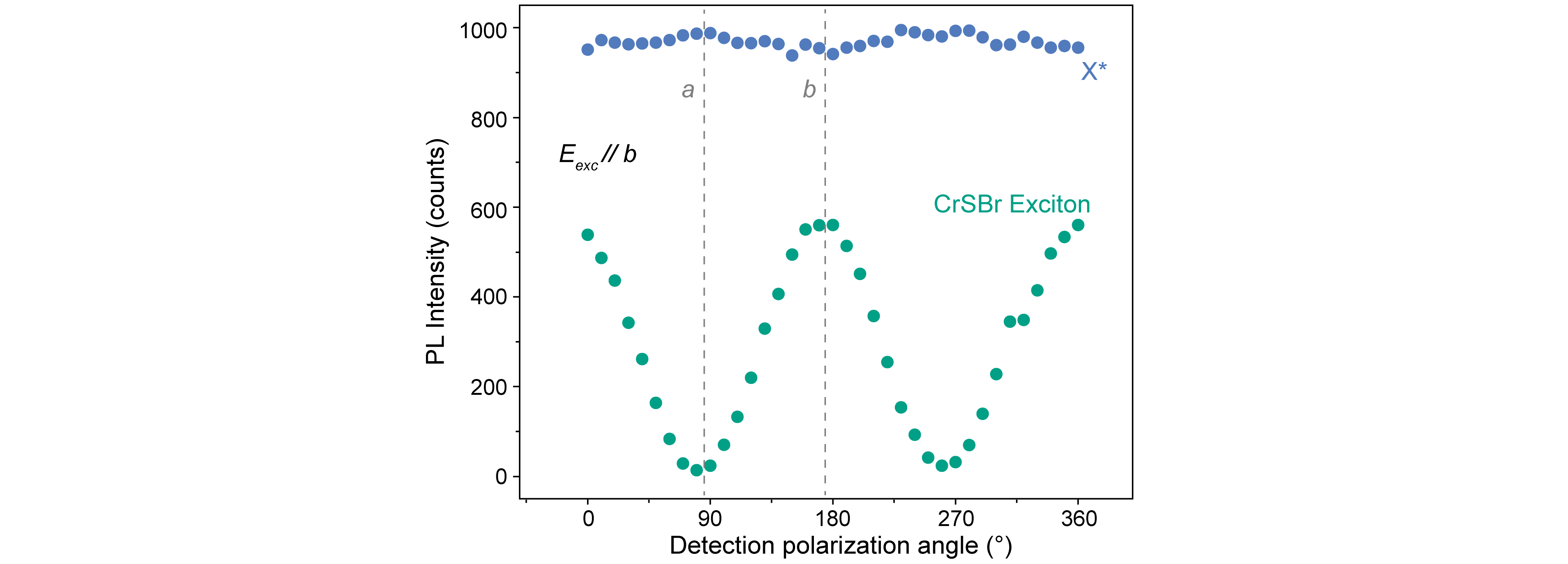}
	\caption{\textbf{PL intensity of X$^*$ and CrSBr exciton as a function of detection polarization angle with excitation polarization along the \textit{b}-axis.} 
	}
	
	\label{baxis}
\end{figure*}

\bibliographystyle{naturemag}
\bibliography{Reference_SI}